\documentclass[prl,twocolumn]{revtex4}
\usepackage{amsmath}
\usepackage{amsfonts}
\usepackage{amssymb}
\usepackage{graphicx}

\begin{document}

\author{Paul Maragakis, Martin Spichty and Martin Karplus}

\affiliation{Department of Chemistry and Chemical Biology, 
Harvard University, 02138 Cambridge MA}
\affiliation{Laboratoire de Chimie Biophysique, 
Institut de Science et
    d'Ing\'enierie Supramol\'eculaires,
Universit\'e Louis Pasteur, 
BP 70028,
8 all\'ee Gaspard Monge, 
F-67083 Strasbourg Cedex, 
France
}

\title{Optimal estimates of free energies from multi-state nonequilibrium work data}

\begin{abstract}
We derive the optimal estimates of the free energies of an
arbitrary number of thermodynamic states from nonequilibrium
work measurements; the work data are collected from forward
and reverse switching processes and obey a fluctuation theorem.
The maximum likelihood formulation properly reweights all pathways
contributing to a free energy difference, and is directly
applicable to simulations and experiments. We demonstrate dramatic gains 
in efficiency by combining the analysis with parallel tempering 
simulations for alchemical mutations of model amino acids.
\end{abstract}
\maketitle

{\em Introduction.---}
Free energy simulations play
an ever increasing role in studies
of condensed matter, particularly
those concerned with problems in 
molecular biophysics.
With the advent of accurate
force fields, 
increases in computer power,
and continuous 
methodological advances,
"free energy simulations [have] 
come of age."~\cite{simonson2002}
Due to
the flexibility of the design 
and control of the simulations
at the
atomistic level, they can provide
information not available
from experiment.
In parallel developments,
a new generation of
single molecule 
studies~\cite{collin2005,cecconi2005}
are using 
advances in 
theory~\cite{bennett1976,jarzynski1997,crooks1999,crooks2000,hummer2001a}
to extract free energy 
information from the 
experiments.  A recent 
example is a single
molecule pulling
experiment~\cite{collin2005}
that determined the folding
free energies of RNA 
strands.
The measured unfolding and
refolding non-equilibrium work 
data~\cite{crooks1999} 
were analyzed with a 
generalization of Bennett's 
acceptance ratio
method~\cite{bennett1976}
to finite switching~\cite{crooks2000}.
The Bennett method
and its generalization 
were shown recently to
provide the maximum
likelihood estimator of the
free energy difference, given
a set of work data between
two states~\cite{shirts2003}.
Surprisingly, the Bennett method,
which dates from 1976,
has rarely been used in computations
of free energy differences
in biological systems; calculations
have been based primarily on the
exponential difference formula of
Zwanzig~\cite{zwanzig1954} or the
thermodynamic integration approach of 
Kirkwood~\cite{kirkwood1935}.

Maximum 
likelihood estimators, under
very general and verifiable conditions,
are asymptotically consistent and efficient estimators~\cite{lehmann1998}; i.e.\  they
provide the smallest variance of any
unbiased estimate of the parameters
underlying the distribution of a large
set of sampled data.
In this letter we provide 
the maximum likelihood estimator
of free energy differences
from  nonequilibrium work data sets
for {\em multiple} states
and verify that it is asymptotically 
unbiased and efficient.
We also show that the method
can be 
used to combine 
all of the sampled data 
from parallel tempering 
simulations~\cite{geyer1991,hansmann1997} 
in an optimum way to
obtain free energy differences.
The analysis is directly applicable 
to present-day
simulations of free energy differences 
of interest in molecular biophysics;
e.g.\  the properties of
mutants or the study of multi-ligand 
binding.
In what 
follows we first derive the 
method 
and then demonstrate 
its utility
in the analysis of 
parallel tempering simulations
of alchemical mutations of 
model amino acids.

{\em Derivation.---}
Consider a set of $N$ 
thermodynamic states, which 
correspond to
systems with different Hamiltonian,
possibly sampled at 
different external conditions
(e.g.\ variation of the temperature).  
A switch from one state 
to another can be produced either
by an instantaneous (sudden) 
change of the 
Hamiltonian and the external conditions with
the two systems in the same microstate~\cite{zwanzig1954,bennett1976},
or a sequence of gradual
changes that lead from the 
Hamiltonian and external conditions 
of the initial state to the
final state~\cite{kirkwood1935,jarzynski1997,crooks1999}.

Consider the pair of thermodynamic states $i$ and $j$ and
a forward and reverse switch,
such that a fluctuation 
theorem~\cite{bennett1976,crooks1999,evans2002}
of the following form
holds for the 
probability distributions of 
a path-dependent quantity $W_{ij}$ 
that is odd under 
path reversal ($W_{ji}=-W_{ij}$):
\begin{equation}
p( W_{ij} | {i \rightarrow j} ) \mathrm{e}^{- W_{ij}}
=
p(W_{ij} | {j \rightarrow i} ) \mathrm{e}^{- A_{ij}},
\label{eq:FT}
\end{equation}
with $p(W_{ij}  | {i \rightarrow j} )$ the
probability of 
measuring $W_{ij}$ along
a path sampled in the switch
from $i$ to $j$, and
$A_{ij}$ an appropriate 
state-dependent quantity;
$W_{ij}$ can be thought of as a 
generalization of the work
and $A_{ij}$ as 
the free energy difference.  
In the instantaneous 
switch case, $W_{ij}$, 
and $A_{ij}$ are:
\begin{eqnarray}
W_{ij}(x) &=& \beta_j E_j (x) - \beta_i E_i (x),
\label{eq:W} \nonumber
\\
A_{ij} &=&  
\ln{Z_i} - \ln{Z_j},
\label{eq:Q}
\end{eqnarray}
with $x$ a microstate (coordinates
and momenta) sampled at equilibrium in 
the initial ensemble, $E_j$ the energy of
state $j$, $\beta_j$ the inverse temperature of 
bath $j$, 
$\ln{Z_j}$ the logarithm of the partition
function of state $j$; with these 
definitions Eq.~\ref{eq:FT} 
holds, as shown in Ref~\cite{bennett1976}.
In the gradual switch case performed
in contact with a constant temperature 
heat bath, 
$W_{ij}(x)= \beta W$, with $W$
the work along the pathway, and
$A_{ij} = \beta \Delta F_{ij}$, with 
$\Delta F_{ij}$ the Helmholtz free
energy difference between states
$i$ and $j$.
More general situations for 
which Eq.~\ref{eq:FT} holds are
discussed in Ref.~\cite{crooks1999,evans2002}.

We follow the analysis 
of Shirts {\em et al}~\cite{shirts2003}
(see also Ref.~\cite{anderson1972})
to obtain the conditional probability 
that a work value $W_{ij}$
along a path between states $i$ and 
$j$ resulted from a sampling of a forward 
($i \rightarrow j$) switching process.  
From Bayes theorem~\cite{bayes1764}, 
the ratio of probabilities of 
the forward to the backward 
directions of the switch 
given the work 
value and the end states $i$, $j$ is:
\begin{eqnarray}
\frac{ p( i \rightarrow j | W_{ij}) }{ p( j \rightarrow i | W_{ij})}
= \frac{p(W_{ij} | i \rightarrow j) p(i \rightarrow j)}{p(W_{ij} | j \rightarrow i) p(j \rightarrow i)},
\label{eq:ratio}
\end{eqnarray}
with $p(i \rightarrow j)$ the probability that the path
between states $i$ and $j$ was sampled in the 
direction from $i$ to $j$; for notational simplicity
we omit the explicit dependence of all probabilities
on the given pair of thermodynamic states $\left\{i,j\right\}$
and on $A_{ij}$.  
A given work value
between the states $i$ and $j$ 
can be sampled in 
either the direction $i \rightarrow j$ or the
direction $j \rightarrow i$, so that the numerator 
and denominator of the left hand side of Eq.~\ref{eq:ratio} sum to 1.
Following Ref.~\cite{shirts2003}, 
Eq.~\ref{eq:ratio} and Eq.~\ref{eq:FT}
can be used to obtain:
\begin{eqnarray}
p( i \rightarrow j | W_{ij}) &=& f( -W_{ij} + A_{ij} + M_{ij} ),
\label{eq:AR}
\\
M_{ij} &=& \ln{n_{ij}^{\mathrm{tot}}} - \ln{n_{ji}^{\mathrm{tot}}},
\label{eq:M}
\end{eqnarray}
with $f$ the Fermi function $f(x) = 1/(1+\mathrm{e}^x)$, and $n_{ij}^{\mathrm{tot}}$
the total number of uncorrelated 
work data in the
direction $i \rightarrow j$. 
For the purpose of the maximum likelihood
estimate of $\ln{Z_i}$, the
ratio $p(i \rightarrow j)/p(j \rightarrow i)$
can be substituted with 
$n_{ij}^\mathrm{tot}/n_{ji}^\mathrm{tot}$
without loss of rigor~\cite{anderson1972,shirts2003}.
Eq.~\ref{eq:AR} resembles
Bennett's acceptance ratio~\cite{bennett1976}
of a 
switch move in a 
simultaneous Monte Carlo sampling
of $i$ and $j$ that minimizes
the variance of their free energy difference.

We can now write the joint likelihood 
$p$ of observing 
forward switches from
all states $i$ to every
other state $j$ given
the work data between
these states as:
\begin{equation}
p = \prod_{i} \prod_{j \neq i} \prod_{n_{ij}} 
f( -W_{ij,n_{ij}} + A_{ij} + M_{ij} ),
\label{eq:p}
\end{equation}
with $W_{ij,n_{ij}}$ the work values 
sampled along the $n_{ij}$ paths 
for the
switches $i \rightarrow j$.
This equation, which follows
from Eq.~\ref{eq:AR} for 
independent work data, 
also gives 
the probabilities of partitioning 
the work data for all pairs of states
into those resulting from 
forward and from reverse switches,
since the reverse process with
$i>j$ corresponds 
to a forward process with $j>i$.

We now determine
the set of $\ln Z_i$ 
that maximizes the probability in Eq.~\ref{eq:p},
or, equivalently the logarithm of this probability.
This is the central result of the 
present development.
As can be seen from
Eq.~\ref{eq:Q}, 
Eq.~\ref{eq:p} remains invariant 
when we  
multiply every partition
function by the same constant.  We can
thus fix one of the $\ln{Z_i}$ 
(e.g. set $\ln{Z_1}$ to zero) and maximize
the logarithm of Eq.~\ref{eq:p} with 
respect to the remaining ones.
Since the derivatives of the objective
function are available in closed
form to all orders, we can use
the Newton-Raphson method to
search for a stationary point.  
Using the properties of the Fermi function:
$\partial{\ln{f(x)}}/\partial{x} = - f(-x)$ and
$f'(x)=-1/(2+2 \cosh{x})$,
we obtain the first and second
derivatives of $\ln{p}$:
\begin{eqnarray}
F_{a} &=&
\frac{\partial{\ln{p}}}{\partial{\ln{Z_a}}}
=
\sum \limits_{ij}
q_{ij}^a
s_{ij},
\label{eq:der}
\\
s_{ij}
&=&
\sum \limits_{n_{ij}}
f( W_{ij,n_{ij}} - A_{ij} - M_{ij} ),
\nonumber
\\
q_{ij}^a 
&=&
-\frac{\partial{A_{ij}}}{\partial \ln{Z_a}}
=
\delta_{j,a}-\delta_{i,a},
\nonumber
\\
H_{ab}
&=&
\frac{\partial^2{\ln{p}}}{\partial{\ln{Z_a}}\partial{\ln{Z_b}}}
=
\sum \limits_{ij}
q_{ij}^a
q_{ij}^b
t_{ij},
\label{eq:hes}
\\
t_{ij}
&=&
\sum \limits_{n_{ij}}
\frac{-1}{2+2
\cosh{( W_{ij,n_{ij}} - A_{ij} - M_{ij} )} },
\nonumber
\end{eqnarray}
with $a$, $b$ the indexes of the states
with respect to which we obtain the 
derivatives and
$\delta_{i,a}$ the Kronecker delta symbol.
The full set of components of the forces, $F_a$
(Eq.~\ref{eq:der}), and of the Hessian, $H_{ab}$
(Eq.~\ref{eq:hes}),
can be efficiently calculated in a single scan through
all pairs of states due to the sparseness
of the array $\{q_{ij}^a\}$.  Furthermore,
the Hessian of Eq.~\ref{eq:hes} is 
always negative definite;
i.e.\ 
direct substitution of 
$q_{ij}^a$, $q_{ij}^b$ in Eq.~\ref{eq:hes}
gives that for an arbitrary nonzero vector 
$y$ of ${\cal R}^N$ the product
$y^\mathrm{T} H y 
= \sum \limits_{ij} t_{ij} \left(y_i - y_j\right)^2$.
This result is strictly smaller than zero
if at least one of the $(y_i-y_j)$ is different
from zero.  The latter is always the case if we
fix one of the components of $y$ to zero, 
which is possible given the invariance
of Eq.~\ref{eq:p} to rescaling of all the
partition functions.
This means that the logarithm of 
Eq.~\ref{eq:p} only has a single
stationary point, which is a maximum.
The maximum of the
probability is obtained efficiently by 
iteration of the Newton-Raphson method:
\begin{eqnarray}
\left(\ln{Z_a}\right)_{n+1} &=& 
\left(\ln{Z_a}\right)_n 
-
\gamma
\sum_b \left({H^{-1}_{ab}}\right)_n 
\left({F_b}\right)_n,
\label{eq:Newton}
\end{eqnarray}
with $\left( \ln{Z_a} \right)_n$ the values
of the partition functions at iteration $n$
and $\gamma$ a scaling factor 
that limits the maximum step size and
increases the radius of convergence.

The limit of the sequence of Eq.~\ref{eq:Newton}
gives the maximum likelihood estimate
of the logarithms of the partition functions of all
the systems and thus the maximum 
likelihood estimate of their free energies.
The maximum likelihood estimate is 
asymptotically unbiased with the 
constraint that the free energies of 
the system cannot become
infinite~\cite{lehmann1998}
(which is always the case
in practice).  
Furthermore, the current estimator
asymptotically has the minimum 
variance since 
the third order derivatives
of the log likelihood are finite 
for any values of the work or the
parameters~\cite{lehmann1998}.  
Thus, for the limit
of a large data set the 
present analysis 
gives the optimal
asymptotically unbiased 
estimate of the free energies.

For $N=2$ systems the
equation $F_a=0$ formally equals that
of the Bennett's acceptance ratio method
and thus, as Bennett 
showed~\cite{bennett1976} it converges
asymptotically to the free
energy perturbation method~\cite{zwanzig1954} 
(or the Jarzynski 
equality~\cite{jarzynski1997}) in the limit of 
equilibrium sampling from only one system.

\begin{figure}
  \begin{center}
 \includegraphics[width=0.95 \columnwidth]{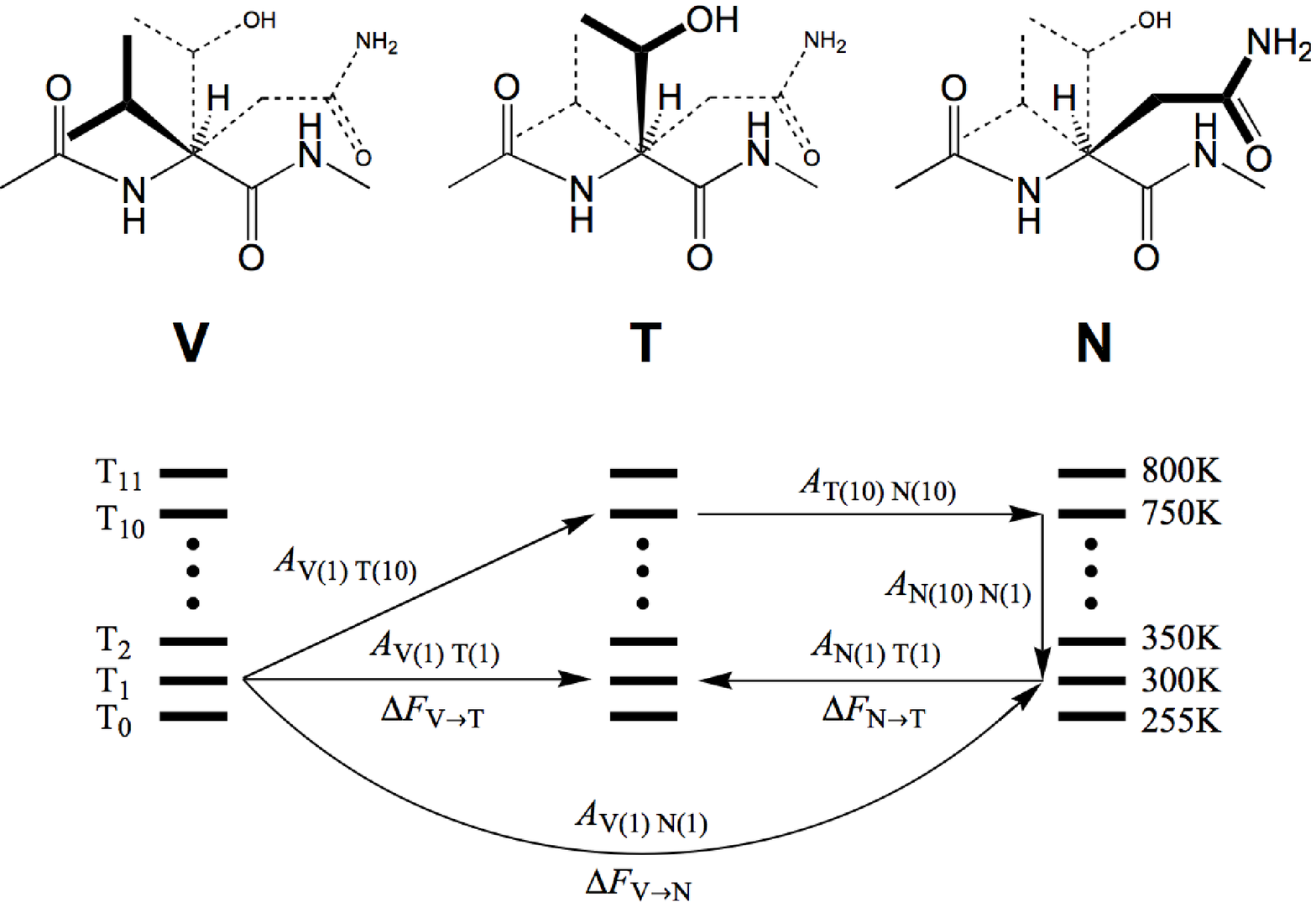}
  \end{center}
  \caption{Sampled thermodynamic states 
for capped amino acids 
valine ($V$), threonine ($T$), and asparagine ($N$). 
The amino acids are modeled with the 
{\sc CHARMM} energy 
function~\cite{brooks1983,mackerell1998}
using a parametric potential:
$V(\lambda_{\mathrm{V}},\lambda_{\mathrm{T}},\lambda_{\mathrm{N}})
 =
 V^{\mathrm{b}}
+
V_{\mathrm{E}\mathrm{E}}^{\mathrm{nb}} 
+
\sum\limits_{X}^{\mathrm{V}, \mathrm{T}, \mathrm{N}  }
{ V_{XX}^{\mathrm{vdw}}}
+
\sum\limits_X^{\mathrm{V}, \mathrm{T}, \mathrm{N}  }
{
\lambda_X (V_{XX}^{\mathrm{elec}} 
				+ V_{X\mathrm{E}}^{\mathrm{nb}})				},
$
with $\lambda_\mathrm{V}$,
$\lambda_\mathrm{T}$, and
$\lambda_\mathrm{N}$ the 
parameters that scale the 
non-bonded and electrostatic
energies of the 
amino acid
of interest,
$V^{b}$ the bonded energy
terms, 
$V_{\mathrm{E}\mathrm{E}}^{\mathrm{nb}}$
the non-bonded energy terms of the
environment $\mathrm{E}$ (backbone),
$V_{XX}^{\mathrm{vdw}}$ and 
$V_{XX}^{\mathrm{elec}}$ 
the van der Waals and electrostatic 
energy terms within 
the side chain of amino acid $X$ 
(from the set $\{\mathrm{V}, \mathrm{T}, 
\mathrm{N}\}$), and
$V_{X\mathrm{E}}^{\mathrm{nb}}$ the
non-bonded energy terms of side 
chain $X$ with the environment.
We implement the
potential with the {\sc BLOCK}
facility~\cite{tidor1990} of 
the {\sc CHARMM} program~\cite{brooks1983}.
The bold side chains
have $\lambda_X=1$, the dashed ones
have $\lambda_X=0$; for example
the state labeled T has
$(\lambda_{\mathrm{V}},
\lambda_{\mathrm{T}},\lambda_{\mathrm{N}})
=(0,1,0)$.
The lower part of the picture shows 
several thermodynamic
states from the parallel tempering simulation.  
The arrows show a small subset of all the 
possible switching pathways
that contribute to the evaluation of the 300 K 
free energy differences.}
  \label{fig:VTN}
\end{figure}

{\em Example.---}
We illustrate the method by 
applying it to the 
analysis of three parallel tempering 
simulations~\cite{geyer1991,hansmann1997}
corresponding
to alchemical mutations in vacuum between 
pairs of the capped amino acids~\cite{brooks1983} 
valine (V),
threonine (T), and asparagine (N) 
as shown in Fig.~\ref{fig:VTN}; 
capped amino acids are 
widely used models in biophysical
studies.
Here, we determine the room temperature 
free energy differences 
$\Delta F_{\mathrm{V} \rightarrow \mathrm{T}}$, 
and 
$\Delta F_{\mathrm{V} \rightarrow \mathrm{N}}$
and their variances as function of
increasing the number of pathways to
demonstrate the power of 
the current approach.

We sample the canonical ensembles
of each amino acid
using a molecular-dynamics--based parallel 
tempering algorithm~\cite{geyer1991,hansmann1997}
that readily equilibrates each 
state despite the high rotational
barriers of the $\chi_1$ 
angles~\cite{straatsma1991,bartels1997}.
Parallel tempering is 
performed with 12 heat baths 
$T_0 = $255 K, $T_1 = $300 K,
$T_{n+1}=300+(n*50)$ K, $n \in \{1, 2, \ldots, 10\}$;
the temperatures are controlled
via Langevin dynamics with a friction
coefficient of 40 ps$^{-1}$.  The 
integration time step is 2 fs; {\sc SHAKE}
constraints~\cite{ryckaert1977} are applied for 
bonds involving hydrogen atoms.   
Every 10 ps we attempt to swap either
all the baths $(T_{2i}, T_{2i+1})$, or 
all the
baths $(T_{2i-1}, T_{2i})$, with 
$i$ integer.  The length 
of each simulation is 360 ns 
(36000 swap attempts).  
We save snapshots 
every 2 ps and obtain 
equilibrated ensembles
of 160,000 snapshots 
for each
heat bath taken from the last 
320 ns of the simulation.

\begin{figure}
  \begin{center}
  \includegraphics{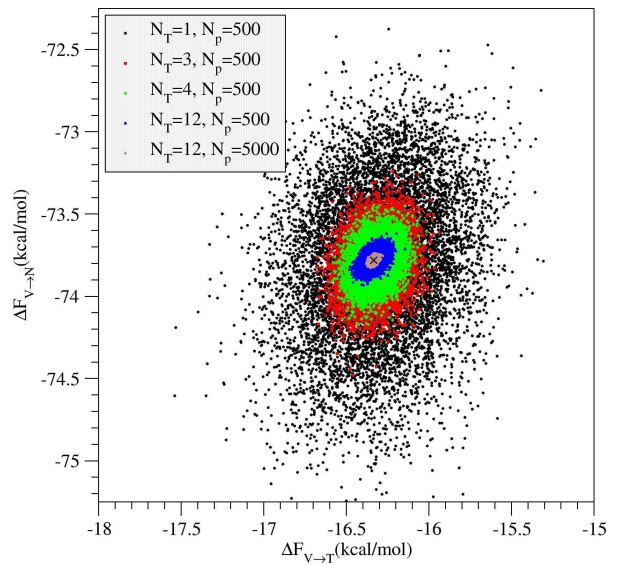}
  \end{center}
  \caption{Scatterplots of the 300 K free
  energy differences for the V$\rightarrow$T and 
  V$\rightarrow$N transitions (see Fig.~\ref{fig:VTN}).  
  Each scatterplot 
  is build from 10,000
estimates of the free energy differences
using random subsets of the work data.
These subsets have length $N_\mathrm{p}$ 
for all possible pairs of the 
$3 \times N_\mathrm{T}$ states of the 
three dipeptides and $N_\mathrm{T}$ 
heat baths.
  The $N_\mathrm{T}=1$
  set contains only the 300K ensembles 
  of V, T, and N; the sets  with more than 
  one temperature contain the ensembles 
  of the $N_\mathrm{T}$ lowest 
  temperature heat baths.  The symbol (X)
  marks the centroid of the set
  with the least scatter.  }
  \label{fig:bullseye}
\end{figure}

Fig.~\ref{fig:bullseye} shows several 
scatterplots of the simultaneous 
estimates of the 300 K
free energy differences
from equilibrium samples 
at $N_\mathrm{T}$ temperatures, 
using $N_\mathrm{p}$
data points for each transition.
The centroid of the $N_\mathrm{T}$=12, 
$N_\mathrm{p}=5000$ set is 
close to the centroid of each scatterplot,
which shows that the estimate is already
unbiased for $N_\mathrm{p}=500$.
The estimates of the free energy  
at 300 K from 
the coordinates of this centroid are
$\Delta F_{\mathrm{V}\rightarrow \mathrm{T}} 
= -16.33 \pm 0.01$ kcal/mol and
$\Delta F_{\mathrm{V}\rightarrow \mathrm{N}} 
= -73.78 \pm 0.01$ kcal/mol; the values
of the free energy are lower for the larger
groups (see Fig.~\ref{fig:VTN}),
essentially due to their more negative energies.

\begin{table}
\begin{tabular}{ccccccc}
$N_{\mathrm{s}}$ & $N_{\mathrm{T}}$ & $N_{\mathrm{p}}$ & 
 $\left< \Delta F_{\mathrm{V}\rightarrow \mathrm{T}} \right>$ 
 & $\sigma_{{\mathrm{V}\rightarrow \mathrm{T}}}$  &
  $\left< \Delta F_{\mathrm{V}\rightarrow \mathrm{N}} \right>$ 
 & $\sigma_{{\mathrm{V}\rightarrow \mathrm{N}}}$  \\
 \hline 
1\footnotemark[1]  & 1 & 500 & -14.22  & 0.85 & -70.68  &0.86\\
1\footnotemark[1] & 1 & $1.6 \times 10^5$ & -16.86  & & -71.72  &  \\
 1\footnotemark[2]  & 1 & 500 &  -18.25 & 0.85 & -76.59  & 1.23 \\
 1\footnotemark[2] & 1 & $1.6 \times 10^5$ & -17.426  & & -74.90  &\\
 2\footnotemark[3]  & 1 & 500 & -16.23  & 0.60 & -73.63  & 0.75\\
2\footnotemark[3] & 1 & $1.6 \times 10^5$ & -17.14  & & -73.31  &\\
 2\footnotemark[4]  & 1 & 500 & -16.31  & 0.31 & -73.81  & 0.51\\
 2\footnotemark[4]  & 1 & 5000 & -16.30  & 0.10 & -73.78 & 0.16 \\
 2\footnotemark[4]  & 1 & $1.6 \times 10^5$ & -16.30  & (0.02) & -73.78  & (0.04) \\
 3\footnotemark[5]  & 1 & 500 & -16.31  & 0.30  &  -73.80 & 0.44 \\
  3\footnotemark[5]  & 3 & 500 & -16.32  & 0.11 &  -73.79 & 0.16\\
  3\footnotemark[5]  & 4 & 500 & -16.32  & 0.08 &  -73.79 & 0.10\\
  3\footnotemark[5]  & 8 & 500 & -16.33  & 0.04 &  -73.78 & 0.05\\
  3\footnotemark[5]  & 12 & 500 & -16.33  & 0.04 &  -73.78 & 0.04\\
  3\footnotemark[5]  & 1 & 5000 & -16.30  & 0.09 &  -73.78 & 0.13\\
  3\footnotemark[5]  & 3 & 5000 & -16.32  & 0.03 &  -73.79 & 0.05\\
  3\footnotemark[5]  & 4 & 5000 & -16.32  & 0.02 &  -73.79 & 0.03\\
    3\footnotemark[5]  & 8,12 & 5000 & -16.33  & 0.01 &  -73.78 & 0.01\\
\footnotetext[1]{Free energy perturbation,initial; $ ^b$Free energy perturbation,final; $ ^c$Free energy perturbation,average; $ ^d$Bennett's acceptance ratio; $ ^e$Multi-state acceptance ratio.}
\end{tabular}
\caption{\label{table:convergence}
The 300 K free energy estimates in kcal/mol
from equilibrium samples of $N_\mathrm{s}$ peptides
at $N_\mathrm{T}$ temperatures, using $N_\mathrm{p}$
data points for each transition. The standard deviations 
$\sigma$ 
were calculated from 10,000 random 
samples; 
those in parentheses are 
analytical estimates
from one sample~\cite{bennett1976}.}
\end{table}

A summary of the scatterplots of Fig.~\ref{fig:bullseye}
and of other types of analysis of the sampled data
is presented in Table~\ref{table:convergence}.
The free energy perturbation 
method~\cite{zwanzig1954} 
is systematically biased for the
small samples~\cite{wood1991}
and gives incorrect
results even when the complete 
data set is used.
The average of the forward 
and backward data~\cite{wood1991} 
is somewhat better but still 
considerably less
accurate than the Bennett 
acceptance ratio analysis
of the same data set. 
Inclusion of more pathways
in the multi-state acceptance
ratio analysis improves the
statistics and keeps the 
estimate consistent.
The striking feature of this table is the scaling
of the multi-state acceptance ratio 
standard deviation $\sigma$; it is
approximately proportional to 
$N_\mathrm{T}^{-1} N_\mathrm{p}^{-\frac{1}{2}}$
for a wide range of temperatures.

To estimate the efficiency
of this method, we used random 
subsets of the microstates,
instead of the random 
subsets of the work used for
Fig.~\ref{fig:bullseye} and 
Table~\ref{table:convergence}.
We picked 500 random 
structures from each 
thermodynamic state
to create the arrays
of the work for all pathways;
these work data are
correlated.  The estimate of 
$\Delta F_{\mathrm{V} \rightarrow \mathrm{N} }$
from 10,000 repetitions of 
this procedure is:
$\Delta F_{\mathrm{V} \rightarrow \mathrm{N}}= -73.78 \pm 0.10$ kcal/mol.
This corresponds to a reduction
of the standard deviation by a factor of 
5.1 compared to that of the 
Bennett acceptance ratio between
V and N at 300 K shown in 
Table~\ref{table:convergence}
(entry $2^\mathrm{d}$, 1, 500);
thus, the analysis of 1 ns of total
simulation obtains the same 
accuracy as 26 ns analyzed
with the Bennett method between
the two end states.

{\em Conclusions.---}
We have presented the
asymptotically
optimal way to obtain the free energy
differences from samples of 
the work data between multiple 
states.  The resulting 
multi-state acceptance ratio 
method is numerically efficient 
and stable.
We have demonstrated the
applicability of this approach to 
the analysis  of parallel tempering
simulations and have shown
that it provides 
estimates of the resulting
free energy differences 
that are more precise 
and accurate than those 
those from the Bennett
method between two states.  
We are applying the approach
to a range of problems
(e.g.\  relative 
solvation free energies of 
amino acids). 
The 
method can be used also
to enhance multi-ligand binding
simulations or the analysis of 
experimental work data~\cite{collin2005,cecconi2005}, 
as an extension of the approach of 
Hummer and Szabo~\cite{hummer2001a}.

We thank Sergei Krivov for discussions
and Leonidas Tsetseris for feedback on
an early draft of the manuscript.
Some of the computations were done 
at the Crimson Grid cluster in Harvard.
The research at Harvard was supported 
in part by a grant from the National 
Institutes of Health.  PM acknowledges 
support by the Marie Curie European
Fellowship grant number 
MEIF-CT-2003-501953.
MS acknowledges support from
the {\sc CHARMM} Development 
Project.

\bibliography{ptr-short}

\begin{thebibliography}{24}
\expandafter\ifx\csname natexlab\endcsname\relax\def\natexlab#1{#1}\fi
\expandafter\ifx\csname bibnamefont\endcsname\relax
  \def\bibnamefont#1{#1}\fi
\expandafter\ifx\csname bibfnamefont\endcsname\relax
  \def\bibfnamefont#1{#1}\fi
\expandafter\ifx\csname citenamefont\endcsname\relax
  \def\citenamefont#1{#1}\fi
\expandafter\ifx\csname url\endcsname\relax
  \def\url#1{\texttt{#1}}\fi
\expandafter\ifx\csname urlprefix\endcsname\relax\def\urlprefix{URL }\fi
\providecommand{\bibinfo}[2]{#2}
\providecommand{\eprint}[2][]{\url{#2}}

\bibitem[{\citenamefont{Simonson et~al.}(2002)\citenamefont{Simonson,
  Archontis, and Karplus}}]{simonson2002}
\bibinfo{author}{\bibfnamefont{T.}~\bibnamefont{Simonson}},
  \bibinfo{author}{\bibfnamefont{G.}~\bibnamefont{Archontis}},
  \bibnamefont{and} \bibinfo{author}{\bibfnamefont{M.}~\bibnamefont{Karplus}},
  \bibinfo{journal}{Accounts Chem. Res.} \textbf{\bibinfo{volume}{35}},
  \bibinfo{pages}{430} (\bibinfo{year}{2002}).

\bibitem[{\citenamefont{Collin et~al.}(2005)}]{collin2005}
\bibinfo{author}{\bibfnamefont{D.}~\bibnamefont{Collin}} \bibnamefont{et~al.},
  \bibinfo{journal}{Nature} \textbf{\bibinfo{volume}{437}},
  \bibinfo{pages}{231} (\bibinfo{year}{2005}).

\bibitem[{\citenamefont{Cecconi et~al.}(2005)}]{cecconi2005}
\bibinfo{author}{\bibfnamefont{C.}~\bibnamefont{Cecconi}} \bibnamefont{et~al.},
  \bibinfo{journal}{Science} \textbf{\bibinfo{volume}{309}},
  \bibinfo{pages}{2057} (\bibinfo{year}{2005}).

\bibitem[{\citenamefont{Bennett}(1976)}]{bennett1976}
\bibinfo{author}{\bibfnamefont{C.~H.} \bibnamefont{Bennett}},
  \bibinfo{journal}{J. Comp. Phys.} \textbf{\bibinfo{volume}{22}},
  \bibinfo{pages}{245} (\bibinfo{year}{1976}).

\bibitem[{\citenamefont{Jarzynski}(1997)}]{jarzynski1997}
\bibinfo{author}{\bibfnamefont{C.}~\bibnamefont{Jarzynski}},
  \bibinfo{journal}{Phys. Rev. Lett.} \textbf{\bibinfo{volume}{78}},
  \bibinfo{pages}{2690} (\bibinfo{year}{1997}).

\bibitem[{\citenamefont{Crooks}(1999)}]{crooks1999}
\bibinfo{author}{\bibfnamefont{G.~E.} \bibnamefont{Crooks}},
  \bibinfo{journal}{Phys. Rev. E} \textbf{\bibinfo{volume}{60}},
  \bibinfo{pages}{2721} (\bibinfo{year}{1999}).

\bibitem[{\citenamefont{Crooks}(2000)}]{crooks2000}
\bibinfo{author}{\bibfnamefont{G.~E.} \bibnamefont{Crooks}},
  \bibinfo{journal}{Phys. Rev. E} \textbf{\bibinfo{volume}{61}},
  \bibinfo{pages}{2361} (\bibinfo{year}{2000}).

\bibitem[{\citenamefont{Hummer and Szabo}(2001)}]{hummer2001a}
\bibinfo{author}{\bibfnamefont{G.}~\bibnamefont{Hummer}} \bibnamefont{and}
  \bibinfo{author}{\bibfnamefont{A.}~\bibnamefont{Szabo}},
  \bibinfo{journal}{Proc. Natl. Acad. Sci. USA} \textbf{\bibinfo{volume}{98}},
  \bibinfo{pages}{3658} (\bibinfo{year}{2001}).

\bibitem[{\citenamefont{Shirts et~al.}(2003)}]{shirts2003}
\bibinfo{author}{\bibfnamefont{M.~R.} \bibnamefont{Shirts}}
  \bibnamefont{et~al.}, \bibinfo{journal}{Phys. Rev. Lett.}
  \textbf{\bibinfo{volume}{91}}, \bibinfo{pages}{140601}
  (\bibinfo{year}{2003}).

\bibitem[{\citenamefont{Zwanzig}(1954)}]{zwanzig1954}
\bibinfo{author}{\bibfnamefont{R.~W.} \bibnamefont{Zwanzig}},
  \bibinfo{journal}{J. Chem. Phys.} \textbf{\bibinfo{volume}{22}},
  \bibinfo{pages}{1420} (\bibinfo{year}{1954}).

\bibitem[{\citenamefont{Kirkwood}(1935)}]{kirkwood1935}
\bibinfo{author}{\bibfnamefont{J.~G.} \bibnamefont{Kirkwood}},
  \bibinfo{journal}{J. Chem. Phys.} \textbf{\bibinfo{volume}{3}},
  \bibinfo{pages}{300} (\bibinfo{year}{1935}).

\bibitem[{\citenamefont{Lehmann and Casella}(1998)}]{lehmann1998}
\bibinfo{author}{\bibfnamefont{E.}~\bibnamefont{Lehmann}} \bibnamefont{and}
  \bibinfo{author}{\bibfnamefont{G.}~\bibnamefont{Casella}},
  \emph{\bibinfo{title}{Theory of Point Estimation}}
  (\bibinfo{publisher}{Springer}, \bibinfo{address}{New York},
  \bibinfo{year}{1998}), \bibinfo{edition}{2nd} ed.

\bibitem[{\citenamefont{Geyer}(1991)}]{geyer1991}
\bibinfo{author}{\bibfnamefont{C.~J.} \bibnamefont{Geyer}}
  (\bibinfo{organization}{Interface Foundation}, \bibinfo{address}{Fairfax
  station, VA}, \bibinfo{year}{1991}), vol. \bibinfo{volume}{Proc. 23rd Symp.
  Interface} of \emph{\bibinfo{series}{Computing Science and Statistics}}, pp.
  \bibinfo{pages}{156--163}.

\bibitem[{\citenamefont{Hansmann}(1997)}]{hansmann1997}
\bibinfo{author}{\bibfnamefont{U.~H.~E.} \bibnamefont{Hansmann}},
  \bibinfo{journal}{Chem. Phys. Lett.} \textbf{\bibinfo{volume}{281}},
  \bibinfo{pages}{140} (\bibinfo{year}{1997}).

\bibitem[{\citenamefont{Evans and Searles}(2002)}]{evans2002}
\bibinfo{author}{\bibfnamefont{D.~J.} \bibnamefont{Evans}} \bibnamefont{and}
  \bibinfo{author}{\bibfnamefont{D.~J.} \bibnamefont{Searles}},
  \bibinfo{journal}{Adv. Phys.} \textbf{\bibinfo{volume}{51}},
  \bibinfo{pages}{1529} (\bibinfo{year}{2002}).

\bibitem[{\citenamefont{Anderson}(1972)}]{anderson1972}
\bibinfo{author}{\bibfnamefont{J.~A.} \bibnamefont{Anderson}},
  \bibinfo{journal}{Biometrika} \textbf{\bibinfo{volume}{59}},
  \bibinfo{pages}{19} (\bibinfo{year}{1972}).

\bibitem[{\citenamefont{Bayes}(1763)}]{bayes1764}
\bibinfo{author}{\bibfnamefont{T.}~\bibnamefont{Bayes}}, \bibinfo{journal}{Phil
  Trans R Soc Lond} \textbf{\bibinfo{volume}{53}}, \bibinfo{pages}{370}
  (\bibinfo{year}{1763}).

\bibitem[{\citenamefont{Brooks et~al.}(1983)}]{brooks1983}
\bibinfo{author}{\bibfnamefont{B.~R.} \bibnamefont{Brooks}}
  \bibnamefont{et~al.}, \bibinfo{journal}{J. Comput. Chem.}
  \textbf{\bibinfo{volume}{4}}, \bibinfo{pages}{187} (\bibinfo{year}{1983}).

\bibitem[{\citenamefont{MacKerell et~al.}(1998)}]{mackerell1998}
\bibinfo{author}{\bibfnamefont{A.~D.} \bibnamefont{MacKerell}}
  \bibnamefont{et~al.}, \bibinfo{journal}{J. Phys. Chem. B}
  \textbf{\bibinfo{volume}{102}}, \bibinfo{pages}{3586} (\bibinfo{year}{1998}).

\bibitem[{\citenamefont{Tidor}(1990)}]{tidor1990}
\bibinfo{author}{\bibfnamefont{B.}~\bibnamefont{Tidor}}, Ph.D. thesis,
  \bibinfo{school}{Harvard University} (\bibinfo{year}{1990}).

\bibitem[{\citenamefont{Straatsma and McCammon}(1991)}]{straatsma1991}
\bibinfo{author}{\bibfnamefont{T.}~\bibnamefont{Straatsma}} \bibnamefont{and}
  \bibinfo{author}{\bibfnamefont{J.}~\bibnamefont{McCammon}},
  \bibinfo{journal}{J. Chem. Phys.} \textbf{\bibinfo{volume}{95}},
  \bibinfo{pages}{1175} (\bibinfo{year}{1991}).

\bibitem[{\citenamefont{Bartels and Karplus}(1997)}]{bartels1997}
\bibinfo{author}{\bibfnamefont{C.}~\bibnamefont{Bartels}} \bibnamefont{and}
  \bibinfo{author}{\bibfnamefont{M.}~\bibnamefont{Karplus}},
  \bibinfo{journal}{J. Comput. Chem.} \textbf{\bibinfo{volume}{18}},
  \bibinfo{pages}{1450} (\bibinfo{year}{1997}).

\bibitem[{\citenamefont{Ryckaert et~al.}(1977)\citenamefont{Ryckaert, Ciccotti,
  and Berendsen}}]{ryckaert1977}
\bibinfo{author}{\bibfnamefont{J.~P.} \bibnamefont{Ryckaert}},
  \bibinfo{author}{\bibfnamefont{G.}~\bibnamefont{Ciccotti}}, \bibnamefont{and}
  \bibinfo{author}{\bibfnamefont{H.~J.~C.} \bibnamefont{Berendsen}},
  \bibinfo{journal}{J. Comp. Phys.} \textbf{\bibinfo{volume}{23}},
  \bibinfo{pages}{327} (\bibinfo{year}{1977}).

\bibitem[{\citenamefont{Wood et~al.}(1991)}]{wood1991}
\bibinfo{author}{\bibfnamefont{R.~H.} \bibnamefont{Wood}} \bibnamefont{et~al.},
  \bibinfo{journal}{J. Phys. Chem.} \textbf{\bibinfo{volume}{95}},
  \bibinfo{pages}{6670} (\bibinfo{year}{1991}).

\end{thebibliography}

\end{document}